\documentclass[pre,twocolumn,showpacs,amsmath,amssymb]{revtex4}
\def\figurewidth{\linewidth}

\usepackage{epsfig}

\begin{document}

\title{Nonequilibrium coupled Brownian phase oscillators}
\author{M. Kostur,$^{1,2}$   J. {\L}uczka, $^2$ and L. Schimansky-Geier $^3$}
\address{$^1$ Department of Physics and Astronomy,  The University of Maine, Orono, Maine 04469\\ 
$^2$ Institute of  Physics,  University of Silesia, 
  40-007 Katowice, Poland \\
$^3$ Institute of Physics, Humboldt-University Berlin, 
Invalidenstr. 110, D-10115 Berlin, Germany
} 
\date{\today}

\begin{abstract}
  A model of globally coupled phase oscillators under equilibrium
  (driven by Gaussian white noise) and nonequilibrium (driven by
  symmetric dichotomic fluctuations) is studied.  For the equilibrium
  system, the mean-field state equation takes a simple form and the
  stability of its solution is examined in the full space of order
  parameters.  For the nonequilbrium system, various asymptotic
  regimes are obtained in a closed analytical form. In a general case,
  the corresponding master equations are solved numerically.
  Moreover, the Monte-Carlo simulations of the coupled set of Langevin
  equations of motion is performed.  The phase diagram of the
  nonequilibrium system is presented. For the long time limit, we have
  found four regimes.  Three of them can be obtained from the
  mean-field theory.  One of them, the oscillating regime, cannot be
  predicted by the mean-field method and has been detected in the
  Monte-Carlo numerical experiments.
\end{abstract}
\maketitle

\ \ \ \\


\section{ Introduction}
\label{sec:intro} 
A system of coupled oscillators has been treated as a model system of
collective dynamics that exhibits a plenty of interesting properties
such as equilibrium and nonequilibrium phase transitions, coherence,
synchronization, segregation and clustering phenomena.  It has been
used to study active rotator systems \cite{shin}, electric circuits,
Josephson junction arrays \cite{wies}, charge-density waves
\cite{stro} oscillating chemical reactions \cite{kura}, planar XY spin
models \cite{aren}, networks of complex biological systems such as
nerve and heart cells \cite{win}.

Such a system of N-coupled phase oscillators is determined  by a set of equations 
of motion in the form \cite{mur} 
\begin{eqnarray}  \label{pha}
\dot x_i = \omega_i + f(x_i) +  \sum_{j=1}^{N} K_{ij} G(x_j, x_i) 
+ \eta_i(t), \nonumber \\ i = 1,..., N, 
\end{eqnarray}
where $x_i$ denotes the phase of the $i$th oscillator and $\omega _i$ is its local 
frequency, i.e. its frequency in the absence of the interaction  between the 
oscillators. The local force is represented by the function $f(x)$ and $G(x, y)$ 
 includes the coupling effect between oscillators. The constants $K_{ij}$ are 
the coupling strengths and $\eta_i(t)$ characterizes fluctuations in the system.
In the case of weak coupling, $G(x,y) = G(x-y)$ and $G$ is a periodic function of 
its argument.  The specific model $G(x)= \sin x$ has been intensively studied  and 
in the physical literature it  is known as a Kuramoto model \cite{kura}. 
If $K_{ij}$ are positive 
then the coupling is excitatory (meaning $x_i$ tends to pull 
$x_j$ toward its value). If $K_{ij}$ are negative 
then the coupling is inhibitory (it tends to increase the difference  between 
$x_i$ and $x_j$).  Most of studies of the model focus on the global coupling 
(each  oscillator interacts with all the other  oscillators), 
where all pairs are interacting with uniform strength, $K_{ij} =K/N$. Then 
the mean-field treatment holds exactly when $N \to \infty$. 

In the paper we study a special case of the model (\ref{pha}) when the
fluctuation term represents thermal-equilibrium and nonequilibrium
fluctuations.  The remainder of this paper is organized as follows.
In the next section we analyze an equilibrium system.  It is a model
with thermal fluctuations being Gaussian white noise.  In Sec.
\ref{sec:noneq}, we study a nonequilibrium system by adding the second
fluctuation source, i.e., a zero-mean, exponentially correlated
symmetric two-state Markov process.  It can describe a case when local
frequencies $\omega_i$ of the oscillators fluctuate in time.  In Sec.
\ref{sec:noneq2}, we present the mean-field numerical solutions of a
corresponding master equation and discuss results of the Monte Carlo
simulations of Langevin equations. Finally, in Sec. \ref{sec:summary}
we formulate the main conclusions.

\section{Mean-field equilibrium system}
\label{sec:eq} 
In this section, we  analyze a system of   phase oscillators in contact with 
thermostat of temperature $T$, namely,  
\begin{eqnarray}  \label{pha1}
\dot x_i = - \sin x_i +  \frac{K}{N} \sum_{j=1}^{N} \sin(x_j - x_i) 
+ \Gamma_i(t), \nonumber \\ i = 1,..., N, 
\end{eqnarray}
where thermal-equilibrium fluctuations $ \Gamma_i(t)$  are modeled by 
zero-mean  delta-correlated Gaussian white noise,  
\begin{eqnarray} \label{ther}
\langle \Gamma_i(t) \rangle =0, \quad 
\langle { \Gamma}_i ({ t}){ \Gamma}_j ({ s}) \rangle = 
2 T \delta_{ij} \delta ({ t}-{ s}).
\end{eqnarray}
This model can represent a planar model with anisotropy or external field. 
More general models than (\ref{pha1}) has been analyzed. Nevertheless, we reconsider 
the simplified model (\ref{pha1}) because of two reasons. Firstly, the state equation 
of the system has a simple tractable form. Secondly, a new aspect of 
the stability problem of states is  presented. 
 
Let us rewrite the interaction term in the form \cite{rei1}
\begin{eqnarray}  \label{equ}
\frac{1}{N} \sum_{j=1}^{N} \sin(x_j - x_i) = s \cos x_i - c \sin x_i,   
\end{eqnarray}
where the averages 
\begin{eqnarray}  \label{sc}
s =  \frac{1}{N} \sum_{j=1}^{N} \sin x_j, \quad 
c =  \frac{1}{N} \sum_{j=1}^{N} \cos x_j
\end{eqnarray}
are order parameters for the system (\ref{pha}). 
In the thermodynamical limit, $N \to \infty$, for each oscillator $x_i=x$ 
the mean-field Langevin equation is obtained from 
the system (\ref{pha}) and reads  
\begin{eqnarray}  \label{lan}
\dot x = F(x,s,c) + \Gamma(t), 
\end{eqnarray}
where the effective force $F(x,s,c) = -  V'(x,s,c)$ (the prime denotes a 
differentiation with respect to $x$) and 
the effective potential 
\begin{eqnarray}  \label{ef}
V(x,s,c) = -(1+Kc) \cos x -Ks \sin x. 
\end{eqnarray}
Let us introduce a probability density 
\begin{eqnarray}  \label{pro}
{\hat P}(x, t) = \langle \delta\left(x(t)-x\right)\rangle  
\end{eqnarray}
of the process (\ref{lan}), where  $x(t)$ is a solution of (\ref{lan}) for a 
fixed realization of noise  $\Gamma(t)$  and  $\langle ...\rangle$ denotes an 
average over all realizations of  $\Gamma(t)$. This density is normalized on a 
real axis, 
\begin{eqnarray}  \label{nor}
\int\limits_{-\infty}^{\infty}  {\hat P}(x, t) \;dx = 1
\end{eqnarray}
and obeys the  Fokker-Planck equation 
\begin{eqnarray} \label{fok}
  \frac{\partial{\hat P}(x,t)}{\partial t} = 
 \frac {\partial} {\partial x}  V'(x,s,c) {\hat P}(x,t)  + 
T \frac {\partial ^2}{\partial x^2} {\hat P}(x,t).  
\end{eqnarray}  
The reduced probability density $P(x, t)$ defined by the relation 
\begin{eqnarray}  \label{red}
P(x, t) = \sum\limits_{n=-\infty}^{\infty}{\hat P}(x+2\pi n, t) 
\end{eqnarray}
satisfies the Fokker-Planck equation (\ref{fok}) as well, is periodic 
\begin{eqnarray}  \label{per}
P(x+2\pi n, t) = P(x, t)  \enskip  \mbox{for any integer}\  n 
\end{eqnarray}
and normalized on one period, 
\begin{eqnarray}  \label{nor2}
\int\limits_{x_0}^{x_0+2\pi} P(x, t) \;dx = 1 \enskip  \mbox{for any real}\  x_0.
\end{eqnarray}
The order parameters $s$ and $c$ are determined self-consistently 
from the set of two equations \cite{kim},  
\begin{eqnarray}     \label{ss}
s = \langle \sin x \rangle &=& \int\limits_{-\infty}^{\infty} 
\sin x {\hat P}(x, t)\;dx \nonumber\\
 &=& \int\limits_0^{2\pi} \sin x P(x, t)\;dx 
\equiv g(s,c), \\ 
\label{cc}
c= \langle \cos x \rangle &=& \int\limits_{-\infty}^{\infty} 
\cos x {\hat P}(x, t)\;dx  \nonumber\\
&=& \int\limits_0^{2\pi} \cos x P(x, t)\;dx 
\equiv h(s,c),
\end{eqnarray}
where  ${\hat P}(x,t) \equiv {\hat P}(x,s,c,t)$ and 
$P(x,t) \equiv P(x,s,c,t)$ depend on parameters $s$ and $c$ via the effective 
one-particle potential $V(x,s,c)$ given by (\ref{ef}).

Our concern is the behavior of the system in the limit of long time, $t\to\infty$.
The stationary state is a thermodynamic equilibrium state and the stationary 
solution $P_{st}(x)$  of (\ref{fok}) is a Gibbs distribution,
\begin{eqnarray}    \label{Pst}
P_{st}(x) = N \mbox{e} ^{-V(x,s,c)/T}, \\
N^{-1} =  \int\limits_0^{2\pi} \mbox{e}^{-V(x,s,c)/T} \;dx. 
\end{eqnarray}
It is well known that in the equilibrium state the average angular velocity vanishes 
(the principle of detailed balance holds)
$\langle \dot x \rangle =0$ (see, e.g., \cite{luczA}). Then  
from (\ref{lan}) it follows that  
\begin{eqnarray}   \label{s=0}
s = \langle \sin x \rangle =  0
\end{eqnarray}
in the stationary state 
and only a symmetric state is realized for which  
the effective potential reduces to the simple form  
\begin{eqnarray}  \label{eff}
V(x,s,c)= V(x,0,c) = -(1+Kc) \cos x.
\end{eqnarray}
The form of this potential is the same as for a system of
non-interacting oscillators, $V(x) = -\cos x$. However, the amplitude
$A=1+Kc$ can change.  If $c > 0$ then the coherence effect occurs and
the most probable state is the deterministic state $x=0$. On the other
hand, if $1+Kc < 0$ then the most probable state changes and the new
state is $x=\pi$.

\begin{figure}[htbp]
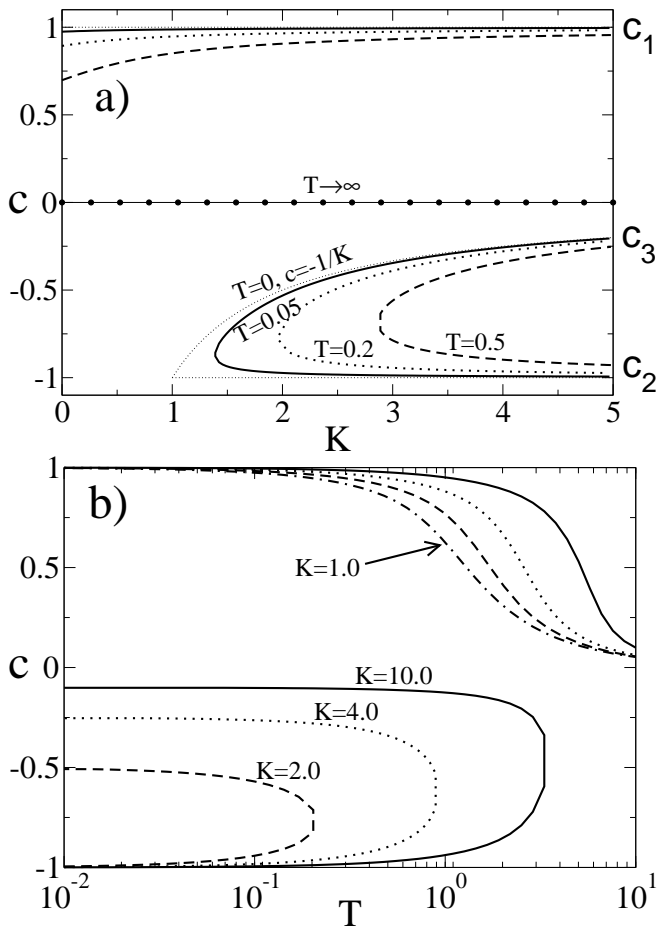

  \begin{center}
     \epsfig{file=./figs/Eq_cK.eps,width=\figurewidth } 
     \epsfig{file=./figs/Eq_cT.eps,width=\figurewidth}  
  \end{center}              
  \caption{ 
    The order parameter $c=\langle \cos x \rangle$ for the system of 
    coupled phase oscillators in 
    equilibrium as a function of the coupling strength $K$ and temperature
    $T$. All data have been obtained as solutions of the implicit
    equation (\ref{c}). The upper plot shows the dependence of $c$ on the
    coupling $K$ for selected temperatures.  Even at $T=0$ the equation
    (\ref{c}) has three solutions $c= (c_1,c_2,c_3) = (1,-1, -1/K)$ for $K>1$ (for
    $K<1$ there exists only one solution $c_1=1$).  The lower plot shows
    the order parameter $c$ as a function of temperature. As it
    could be expected, for large thermal fluctuations, stochastic
    forces overwhelm the potential and the coupling, and the solution
    tends to $c_1=0$ as $T\to \infty$.  The stability analysis shows
    that stable are only solutions with $c > 0$.
 }
\label{fig1}
\end{figure}

\begin{figure}[htbp]
  \begin{center}
     \epsfig{file=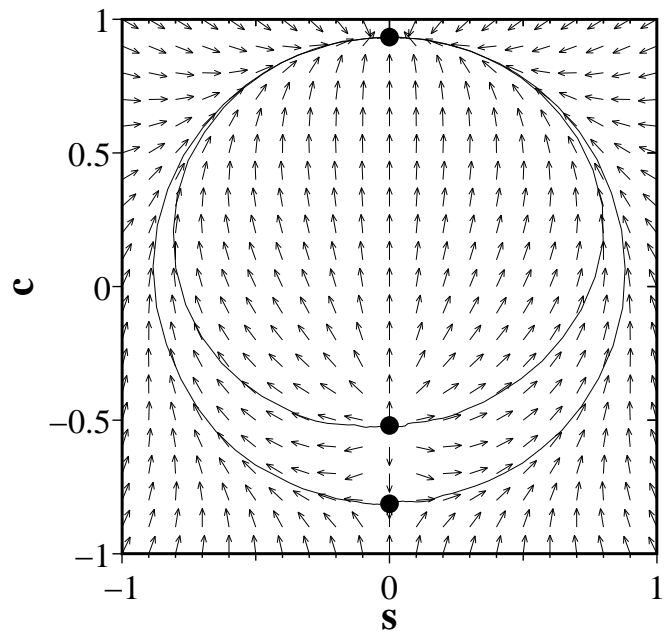,width=\figurewidth} 
  \end{center}              
  \caption{ 
    The plot shows a vector field of the  right hand side of the dynamical system 
    (\ref{sdot})-(\ref{cdot}) for
    $K=3.1$ and $T=0.5$. Three black dots are solutions of the mean-field
    problem, i.e.  stationary points of (\ref{sdot})-(\ref{cdot}). The upper
    solution $c_1$ is a stable node, the middle one $c_3$ is an unstable node and
    the lowest one $c_2$ is a saddle point. Let us notice that stability
    analysis in one dimension (assuming that $s=0$) would lead to a
    false conclusion that the lowest point $c_2$ is stable.  Solid lines are
    a result of the Monte Carlo simulation of $8000$ particles with 
    the initial condition
    set to the  mean-field solutions $c_2$ and $c_3$. In the $(s,c)$-space the system
    evolves along the clockwise or anticlockwise  ``semicircles''  (depending on
    the initial microstate) to the stable node $(0,c_1)$.  }
\label{fig2}
\end{figure}

The order parameter $c$ is determined by the equation
\begin{eqnarray}  \label{c}
c  I_0\left(\frac{1+Kc}{T}\right) = I_1\left(\frac{1+Kc}{T}\right),
\end{eqnarray}
where $I_0(z)$ and $I_1(z)$ are the modified Bessel functions.  This
equation can possess one, two or three solutions (see Fig.
\ref{fig1}).  If the coupling strength $K < 1$ then only one solution
exists for any temperature $T$ of the system (Fig.  \ref{fig1}a).  
  For high temperature, $T>>1$, the upper
branch $c_1$ tends to zero as $c \sim T^{-1}$ (Fig. \ref{fig1}b).  The
opposite asymptotics, when $T \to 0$, can be obtained as well.  In
this case the upper branch $c_1 \to 1$ for any $K > 0$. The lower
branch $c_2 \to -1$ and the middle branch $c_3 \to -1/K$ for $K > 1$
(Fig.  \ref{fig1}a).  
Now, let us study  stability of the stationary solutions. 
The  linear stability analysis
should be performed on the full set of equations of motion for 
average  values  $s$ and $c$,  (\ref{ss})
and (\ref{cc}). Multiplication of (\ref{fok}) by either $\sin x$ or
$\cos x$ and integration over $x$ gives
\begin{eqnarray}
  \label{meanf11}
\dot{s} &=&-(1+K\,c)\,<sc>\,+\,T\,s\,+\,K\,s\,<c^2>,\\ 
 \label{meanf12}
\dot{c} &=& (1+K\,c)\,<s^2>\,-\,T\,c-\,K\,s\,<sc>,  
\end{eqnarray}
and $<...>$ stands for the  averages of products of $\cos x$ and/or
$\sin x$ (e.g. $<sc> = <\sin x \cos x>$). To make the 
system (\ref{meanf11})-(\ref{meanf12})  closed, 
we should write equations of motion for the  unknown statistical moments 
$<sc>$, $<s^2>$ and $<c^2>$. New,  higher-order moments will occur and in 
this way we obtain a hierarchy of infinite number of differential equations 
for moments, which is difficult to handle. Therefore we proceed in 
another way.     Let us notice that 
for $<s^2>\equiv <\sin^2 x>$ one may write
$<s^2> = 1-<\cos^2 x> = 1-<c^2>$.  Additionally, one can
introduce deviations from the mean values and write $<c^2>=c^2+<(\delta
c)^2>$ as well as $<sc>=s\,c+<\delta s\delta c>$. As a  result, one obtains
\begin{eqnarray}   \label{meanf21}
\dot{s}  =  -(c - K\,<(\delta c)^2> - T )\,s \nonumber \\
-(1+K\,c)\,<\delta s \delta c>,   \\ 
 \label{meanf22}
\dot{c}  =  (1+K\,c)\,(1-c^2-<(\delta c)^2> ) - T\,c - K\,s^2\,c \nonumber\\
-K\,s\,<\delta s \delta c>.
\end{eqnarray}

From (\ref{s=0}) we know that $s=0$ is a stationary solution of the 
above equations.  In order 
to obtain  this solution $s=0$ from (\ref{meanf21}),  
the correlator $<\delta s\delta c>$ 
should vanish in the stationary limit, i.e., $<\delta s\delta c> \to 0$ 
as $t \to \infty$.  Insertion of $s=0 $
into the second equation (\ref{meanf22}) with $\dot{c}=0$
yields stationary  solutions for $c$. They are determined by the equation  
\begin{equation} \label{cst}
(1+K c)\,(1-c^2-<(\delta c)^2>) - T c = 0 
\end{equation}
These solutions depend on the unknown 
variance  $<(\delta c)^2>$. In the low temperature limit $T\to 0$,
the variance $<(\delta c)^2> \to 0$ and we recover the solutions 
$c_1=1, c_2=-1$ and $c_3=-1/K$.  In this case, the linear stability analysis  
of (\ref{meanf21})-(\ref{meanf22}) 
shows that the stationary point $(0, c_1)$ is  a  stable node,  the   
 point $(0, c_2)$ is a saddle and the solution $(0, c_3)$ is an 
unstable node. For $T > 0$, 
the stability of solutions remains unchanged. 
Indeed, in our simulations we have confirmed this statement. We have also 
analyzed an auxiliary dynamical system defined by   a set
of two differential equations, namely (cf. (\ref{ss}) and (\ref{cc})),
\begin{eqnarray}     \label{sdot}
\dot s = -s +  g(s,c), \\ 
\label{cdot}
\dot c= -c +  h(s,c).
\end{eqnarray}
The stationary solution of this system is the same as the equilibrium state 
of the system (\ref{lan}). 
In Fig. \ref{fig2} we present a vector field generated by 
the  dynamical system (\ref{sdot})-(\ref{cdot}) and its three 
stationary points 
$(s_i, c_i), i=1,2,3$. One can infer that the 
upper point $(0, c_1)$ is  a  stable node,  the lower   point $(0, c_2)$ is a saddle and the middle point $(0, c_3)$ is an
unstable node 
(the same as for (\ref{meanf21})-(\ref{meanf22})).  
We have also  found unexpectedly that the trajectory of the system 
(\ref{sdot})-(\ref{cdot}) is the same as that obtained from simulations 
of the set of Langevin equations (\ref{pha1}), see  Fig. \ref{fig2}.   
It allows us to formulate  the  conjecture  that 
the hierarchy of infinite number of equations for moments   of the set 
$(\sin x, \cos x)$is equivalent to (\ref{sdot})-(\ref{cdot}). 
Unfortunately, we cannot prove it rigorously.  

\section{Mean-field nonequilibrium system}
\label{sec:noneq} 
Nonequilibrium  systems can be modeled by including a term 
which describes non-thermal and nonequilibrium fluctuations, noise and perturbations. 
There are many possibilities to do this  but 
here we consider a slight modification of the previous model, namely, 
\begin{eqnarray}  \label{pha2}
\dot x_i = - \sin x_i +  \frac{K}{N} \sum_{j=1}^{N} \sin(x_j - x_i) 
+ \Gamma_i(t) +\xi_i(t), \nonumber \\ i = 1,..., N, 
\end{eqnarray}
 The random functions  
 $\xi_i(t)$ represent nonequilibrium fluctuations and are modeled by 
a {\it symmetric} dichotomic Markovian stochastic processes \cite{chem},
\begin{eqnarray} \label{dich}
\xi_i (t)=\{-a,a\},\mbox{ }a\mbox{ }>0, \\  \nonumber
P(-a\rightarrow a)=P(a\rightarrow -a)=\mu ,
\label{war1}
\end{eqnarray}
where $P(-a\rightarrow a)$ is a probability per unit time  of the jump from 
the state $-a$ to the state  $a$. This process is of zero average, 
$\langle \xi_i (t)\rangle =0$, and exponentially correlated, 
\begin{equation}
\label{cor1}
\langle \xi_i (t)\xi_j (s)\rangle = a^2 \delta_{ij} e^{-\left| t-s\right| / \tau },
\end{equation}
 where $\tau=1/2 \mu$ is  correlation time of the  process $\xi_i(t)$. So, it 
 is characterized by two parameters: its
amplitude $a$ (or equivalently the variance $<\xi ^2(t)>=a^2$) 
 and the correlation time $\tau$.

The mean-field Langevin equation takes the form 
\begin{eqnarray}  \label{lan2   }
\dot x = -  V'(x,s,c) + \Gamma(t) + \xi(t)  
\end{eqnarray}
and the corresponding master equations read \cite{van}
\begin{eqnarray} \label{ma1}
 \frac{\partial P_{+}(x,t)}{\partial t} =   
 \frac \partial {\partial x} 
\left[V'(x,s,c)-a\right]P_{+}(x,t)  \nonumber \\ 
+ T \frac{\partial ^2}{\partial x^2}P_{+}(x,t)
- \mu P_{+}(x,t)+\mu P_{-}(x,t)   \\ 
\label{ma2}
 \frac{\partial P_{-}(x,t)}{\partial t} =
  \frac \partial {\partial x}
          \left[V'(x,s,c)+a\right]P_{-}(x,t)  \nonumber \\ 
+ T \frac{\partial ^2}{\partial x^2}P_{-}(x,t)
           +  \mu P_{+}(x,t)-\mu P_{-}(x,t)   
\end{eqnarray} 
where the probability densities 
\begin{eqnarray}  \label{pra}
P_{+}(x,t)\equiv P(x,a,t),\quad P_{-}(x,t)\equiv P(x,-a,t).
\end{eqnarray}
depend on the order parameters  $s$ and $c$, which in turn depend 
self-consistently on the marginal density 
$ P(x,t) = P_{+}(x,t) + P_{-}(x,t)$ 
via the relations (\ref{ss})-(\ref{cc}). 
Eqs (\ref{ma1})-(\ref{ma2}) cannot be solved analytically, even in the stationary 
state. However, in some limiting cases, stationary solutions of them are known, 
e.g., if the correlation time 
$\tau \to \infty$ (the adiabatic limit) or 
 if temperature of the system is zero, $T=0$.

\subsection{Analytical results}
\label{sec:noneq1} 
From the ratchet theory we know that  the stationary 
average  angular velocity is zero, 
$\langle v \rangle = \langle \dot x \rangle = 0$, 
because the potential (\ref{ef}) is symmetric and fluctuations (\ref{dich}) 
are symmetric \cite{luczA}.   
Therefore 
\begin{eqnarray}   \label{s=00}
s = \langle \sin x \rangle =  0
\end{eqnarray}
and $V(x,s,c)$ takes the same form as in the previous case 
(\ref{eff}). In the adiabatic limit, 
the equation determining a stationary state is 
\begin{eqnarray}  \label{c2}
c = \frac{1}{2} \int\limits_0^{2\pi} \cos x \; 
\left[p_{+}(x,c) + p_{-}(x,c) \right] dx,
\end{eqnarray}
where the stationary probability densities 
$p_i(x,c), \left(i=+,-\right)$ read 
\begin{eqnarray}    \label{Pi}
p_i(x,c)  =  \frac{U_i(x,c) \int\limits_x^{x+2\pi} U^{-1}_i(y,c) dy } 
{\int\limits_0^{2\pi} U_i(x,c) \int\limits_x^{x+2\pi} U^{-1}_i(y,c) dy dx} 
\end{eqnarray}
and 
\begin{eqnarray}    \label{Ui}
U_{\pm}(x,c) =  \mbox{e}^{V(x,0,c)/T}  \mbox{e}^{\pm ax/T}.
\end{eqnarray}
%


In the second limit, i.e. when temperature of the system is zero, $T=0$, 
the stationary state is determined by the equation 
\begin{eqnarray}    \label{cT0}
c  =  \frac{\int\limits_{\Omega(c)} \cos x \;  D^{-1}(x,c) \mbox{e}^{-\Psi(x,c)} \;dx}
{\int\limits_{\Omega(c)} D^{-1}(x,c) \mbox{e}^{-\Psi(x,c)} \;dx}.
\end{eqnarray}
where the thermodynamic potential
\begin{eqnarray}    \label{Psi}
\Psi(x,c)   =  \int\limits_0^{x} D^{-1}(y,c) V'(y,0,c) \;dy
\end{eqnarray}
and  the effective diffusion function 
\begin{eqnarray}    \label{D}
D(x,c)  =  \tau \left[a^2-V'(x,0,c)^2\right].
\end{eqnarray}
The integration interval $\Omega(c) = [0, 2\pi]$ iff $D(x,c) > 0$. If in some 
intervals  the  function $D(x,c)$ is negative then 
$\Omega(c) = [x_1, x_2]$, where $x_1$ and $x_2$ are suitable roots of the equation 
 $D(x,c) = 0$ and in the interval $[x_1, x_2]$ the diffusion function
 is positive. 

\begin{figure}[htbp] 
  \begin{center}
     \epsfig{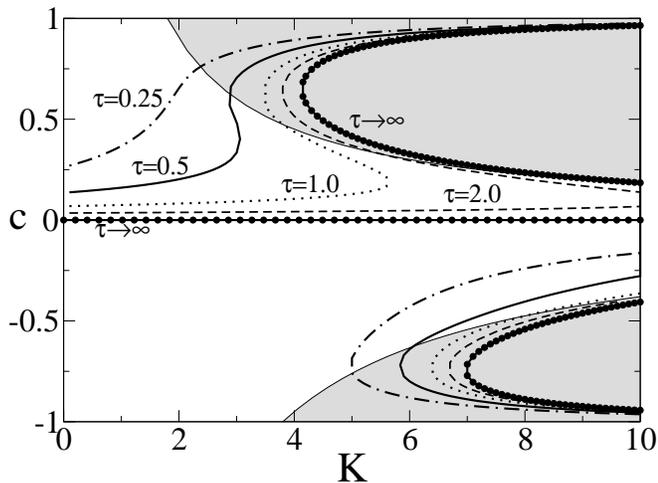} 
  \end{center}              
  \caption{ 
    Solutions of the stationary mean-field problem (\ref{ss}),(\ref{cc}), 
    (\ref{ma1}) and (\ref{ma2}). The temperature
    is zero and the amplitude of dichotomic fluctuations is $a=2.8$. As in
    the equilibrium case, negative solutions are unstable. Stable, and
    observed in Monte-Carlo simulations solutions lie in the upper plane
    $c>0$. Grey regions depict places where the locking condition is met
    i.e. maximum of the effective force overwhelms the amplitude of
    fluctuations. One can notice that for small values of $\tau$, the 
    system starts to behave as an equilibrium one. 
     In the case $\tau \to \infty$ the asymptotic analytical solution 
     is shown. For finite
    $\tau$ numerical results are depicted. In this case the system exhibits
    onset of hysteresis in $c(K)$.  }
\label{fig3}
\end{figure}

The  limiting case $T=0$ and $\tau \to \infty$  is analytically tractable. 
 From the master equations it follows that in this case  the stationary 
state is determined  by the equation 
\begin{eqnarray}    \label{PT0}
\left[a^2-V'(x,0,c)^2\right]P(x)  =  const. 
\end{eqnarray}
In the {\it diffusive regime}, when dichotomic noise activates both forward and 
backward transitions over barriers of the effective potential, the solution of 
(\ref{PT0}) is 
\begin{eqnarray}    \label{diff}
P(x)  =  const. /\left[a^2-V'(x,0,c)^2\right]
\end{eqnarray}
and the only solution of the state equation (\ref{cc}) is $c=0$. 
In the {\it non-diffusive regime}, when dichotomic noise cannot activate  neither  
forward nor  
backward transitions over barriers of the effective potential, the normalized 
solution of  (\ref{PT0}) has the form 
\begin{eqnarray}    \label{delt }
P(x)  = \frac{1}{2}\left[\delta(x-x_1)+\delta(x-x_2) \right]
\end{eqnarray}
where $x_1$ and  $x_2$ are solutions of the equation  
\begin{eqnarray}    \label{sol1}
a^2-V'(x,0,c)^2 = 0. 
\end{eqnarray}
If $1+Kc > 0$ then the state equation is determined by 
\begin{eqnarray}    \label{c4}
 c = \cos\left[\arcsin(a/(1+Kc))\right] 
\end{eqnarray}
This equation can possess two positive roots, $c_1 > c_2 > 0$. 
The solution $c_1$ is stable while $c_2$ is unstable.   
If $1+Kc < 0$ then
\begin{eqnarray}    \label{c5}
 c = - \cos\left[\arcsin(a/(1+Kc)) \right]. 
\end{eqnarray}
This equation can possess two negative  roots which are  unstable.   
It is  depicted in Fig. \ref{fig3}.

\subsection{Numerical methods}
\label{sec:noneq2} 
In a general case  the mean-field problem reduces
to the set of non-linear master equations (\ref{ma1}) and (\ref{ma2})
which has to be solved. Apart of a few previously considered special 
cases which can be
treated analytically, only numerical methods are applicable.  We have
approached the numerical problem of solving (\ref{ma1}) and (\ref{ma2})
as follows.  Conditions of self-consistency have been considered
(\ref{ss}) and (\ref{cc}) as the non-linear minimization problem in
two dimensions on the bounded domain $-1\le s \le 1$ and 
$-1 \le c \le 1$. It has been
handled in a standard way, making use of numerical libraries.
However, each evaluation of functions $g$ and $h$ for given $(s,c)$
requires the knowledge of stationary solution of the system
(\ref{ma1}), (\ref{ma2}) with fixed $s$ and $c$. This is, in turn, a
linear boundary value problem  which can be easily solved with
the help of finite element method (FEM). In the case $T=0$, the
stationary distribution $P(x)$ is given be quadratures but it is very difficult to 
handle it analytically. Moreover, we
found it practically easier to obtain a FEM solution for small enough
$T$ then to estimate, divergent in some cases, triple integrals.

Additionally, in order to verify mean-field results we have performed
Monte-Carlo simulations of the Langevin equations (\ref{pha2}). This,
independent method enabled not only verification of numerical results
but also applicability of the mean-field approach. Because the Monte-Carlo
simulation follows the evolution of microscopic state of the system it
can be considered as a numerical experiment, in contrary to
the mean-field approach which is only an approximation. In general,
Monte-Carlo simulations of globally interacting $N$-particles require
$\simeq N^2$ operations per time step. The special form of the interaction
term $\simeq \sin(x_j-x_i)$, leads to relations (\ref{equ}) and
(\ref{sc}). In the course of simulation the average values $s$ and $c$
need to be evaluated only once per a  simulation step, what reduces the
number of operations per a time step to $\simeq N$. 

\subsection{General case: numerical analysis}
\label{sec:noneq3} 
\begin{figure}[htbp]
  \begin{center}
    \epsfig{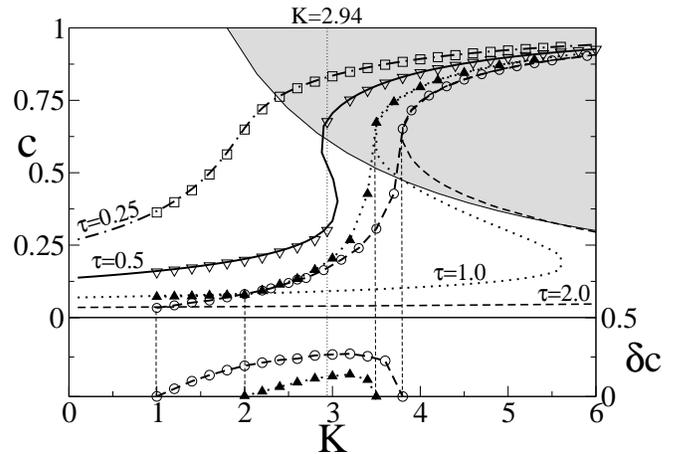}
  \end{center}              
  \caption{
    The same case as in Fig. \ref{fig3}: the comparison of
    mean-field results and Monte-Carlo simulations. For $\tau=0.25$ and
    $\tau=0.5$ there is a perfect agreement of the Monte Carlo and
    mean-field methods. However, for $\tau=1$ and $\tau=2$ temporal
    oscillations of density of particles appear in the Monte-Carlo method.
    Thus the order parameter also performs temporal oscillations.
    An averaged value of those oscillations differs from those coming
    from the mean-field solution. The standard deviation of $c$ (averaged
    in time) is shown in the  lower insert. One can notice that if 
    oscillations disappear ($\delta c=0$) then the simulated values of
    $c$ agree very well with mean-field predictions.}
\label{fig4}
\end{figure}

All numerical mean-field results have been obtained in the stationary
regime.  First we study the zero-temperature case, $T=0$.  
The natural characteristics of the stationary state are statistical moments, 
in particular the first two moments $<x>$ and $<x^2>$.  
 These moments are not good characteristics in the case considered. 
If the system is spatially periodic, then for any spatially periodic 
function $A(x) = A(x+2\pi)$ we can calculate its mean value exploiting 
either the  probability density ${\hat P}(x, t)$ or the reduced probability 
distribution $P(x, t)$ because then the equality 
\begin{eqnarray}     \label{A(x)}
 \langle A(x) \rangle = \int\limits_{-\infty}^{\infty} 
A(x)  {\hat P}(x, t)\;dx  =
\int\limits_0^{2\pi} A(x) P(x, t)\;dx
\end{eqnarray}
holds. It is not a case for 
non-periodic functions and then there is a problem which the distribution 
should be used for calculationg the average value. 
Therefore we consider periodic functions. Here, two natural 
order parameters $s=<\sin x>$ and $c=<\cos x>$ occur which characterize 
the probability distribution in the same way as $<x>$ and $<x^2>$. Indeed,   
the function $\sin x$ is odd like the function $x$ and the function 
$\cos x$ is even like the function $x^2$.  
Because $<\sin x> = 0$, below we analyze $<\cos x>$.   
In Fig.
\ref{fig3} we show the dependence of the order parameter $c=\langle
\cos x\rangle$ on the coupling strength $K$.  One can distinguish two
main regimes: the diffusive (dichotomic noise activates transitions
over barriers of the effective potential (\ref{eff})) and 
non-diffusive or locked
(dichotomic noise cannot activate transitions over barriers of the
effective potential (\ref{eff})).  These two regimes, marked by white and gray
regions in Fig. \ref{fig3}, are separated by two critical lines:
$Kc+1=a$ for positive values of $c$ and $Kc+1=-a$ for negative values
of $c$. For negative $c$, the dependence of $c$ upon $K$ is
qualitatively the same as for the equilibrium system (Fig.
\ref{fig1}). These solutions are unstable and therefore will not be
considered.  Now, let us discuss the positive solutions $c>0$.  They
depend strongly on the correlation time $\tau$ of dichotomic
fluctuations.  For short correlation time, the order parameter $c$
monotonically increases with the growing coupling $K$.  For longer
correlation time, new effects arise: the dependence is discontinuous
and hysteretic. In some domain there are three solutions $c_1 > c_2 >
c_3$.  The solutions $c_1$ and $c_3$ are stable while $c_2$ is
unstable. The hysteresis is bigger and bigger when $\tau$ increases.
The jumping point $K_1$ from the lower to the upper branch tends to
infinity and the jumping point $K_2$ from the upper to the lower
branch tends to a constant value determined by eq. (\ref{c4}). The
upper branch of solutions $c_1(K) \to 1$ and the lower branch $c_3(K)
\to 0$ when $\tau \to \infty$. For $\tau = \infty$, the solutions
split into two branches of three solutions, namely, one $c_3=0$ and two other
determined by (\ref{c4}).  The stationary mean-field solutions have
been verified by the Monte-Carlo simulations. The comparison is
presented in Fig. \ref{fig4}.
Simulations show that the implicit assumption of time-independent
stationarity of the system (when $t\to\infty$) is restricted to some
values of parameters of the model.  Indeed, if the time-independent
stationary state of the system exists then the mean-field solutions
agree with simulations.  In particular, for $\tau = 0.5$ the
hysteresis is observed (see point $K=2.94$ in Fig. \ref{fig4} and Fig.
\ref{fig7}).  However, for longer correlation time $\tau$, temporarily
oscillating steady-states exist for which the probability distribution
$P(x, t)$ is periodic in time. In consequence the order parameters
$s=s(t)$ and $c=c(t)$ are time-periodic and in the limit of long time,
the time-dependent steady-states appear. This is the case when the
mean-field predictions fail, e.g.  the hysteresis is not realizable.
In Fig \ref{fig4} we depicted this phenomenon for $\tau =1, 2$. We
have noticed only monotonic dependence of $c$ upon $K$ (if $c$ is a
periodic function of time, its time average is taken). The quantity
which can characterize the time-independent
stationarity/time-dependent stationarity (i.e. oscillations) of the
long time state is the time-averaged standard deviation $\left(\delta
  c\right)^2 = \langle c^2 \rangle_t -\langle c \rangle_t^2$ of the
order parameter.  We have observed that if $\delta c = 0$ then the
mean-field solutions are correct.  Otherwise, they are incorrect. It
is shown in the lower insert of Fig. \ref{fig4}.

\begin{figure}[htbp]
  \begin{center}
     \epsfig{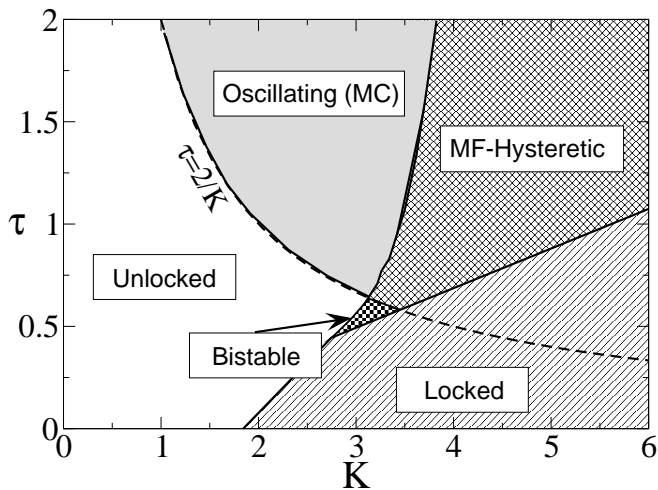} 
  \end{center}              
  \caption{ 
    The phase diagram of  the system with $a=2.8$ and $T=0$. Five
    various regimes are distinguished: unlocked and oscillating,
    mean-field hysteretic, locked  and bistable. 
    The oscillating
    regime has been verified by  Monte-Carlo simulations. All other data
    come from the stationary mean-field problem. The empirical formula
    $\tau=2/K$ surprisingly well fits the left boundary of the
    oscillating region. }
\label{fig5}
\end{figure}

\begin{figure}[htbp]
  \begin{center}
    \epsfig{file=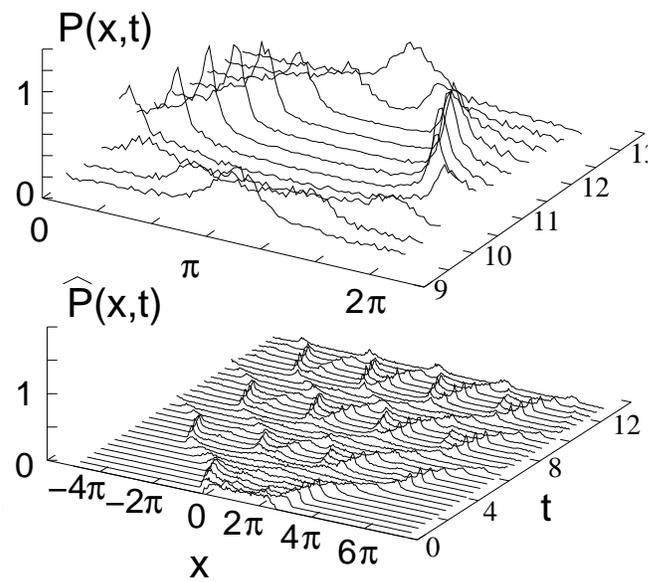,width=\figurewidth}
  \end{center}              
  \caption{  
    Monte Carlo simulations of the system for $a=2.8$, $\tau=1.0$,
    $K=2.94$ and $T=0$. None of the mean-field prediction is realized.
    The only stable solution is a stationary one. In the upper plot
    evolution of the probability density reduced to $x\in (0,2\pi)$ is
    shown. In lower plot the full distribution is presented. A
    starting value was a uniform distribution on $x \in (0,2\pi)$.
}
\label{fig6}
\end{figure}

\begin{figure}[htbp]
  \begin{center}
    \epsfig{file=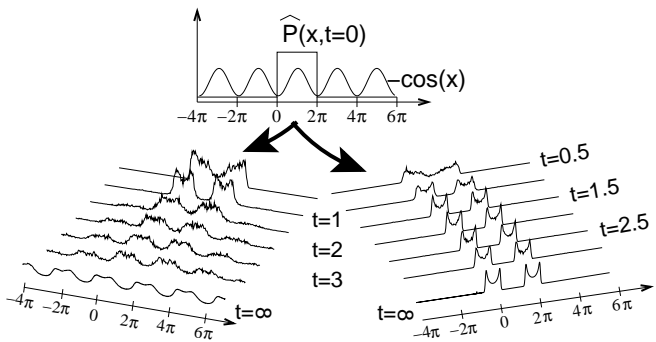,width=\figurewidth}
  \end{center}              
  \caption{ 
    Monte Carlo simulations of the system for $a=2.8$, $\tau=0.5$,
    $K=2.94$ and $T=0$.  The starting point was $8000$ particles
    distributed uniformly on $x\in (0,2\pi)$. The only difference
    between left and right scenarios is the microscopic state:
    individual particles were chosen differently (all macroscopic
    parameters are the same). The left scenario leads to diffusive
    state i.e. $c<(2.8-1)/K$ while the right one leads to locked one
    $c>(2.8-1)/K$. The shape of a stationary, mean-field distribution
    is shown for $t \to \infty$.
}
\label{fig7}
\end{figure}

In Fig. \ref{fig5} we present the phase diagram on the 
$\left(K, \tau\right)$ plane for a fixed amplitude $a=2.8 $ of dichotomic
fluctuations. In the case of non-interacting oscillators, this value
of $a$ corresponds to the diffusive regime.  Roughly speaking, there
are two regions: diffusive when $a > 1+Kc(\tau)$ 
(i.e. dichotomic noise can induce  transitions
over barriers of the effective potential (\ref{eff})) 
and non-diffusive when $a < 1+Kc(\tau)$ (i.e. dichotomic noise cannot induce 
transitions over barriers of the effective potential (\ref{eff})). 
 The diffusive region is divided into two
parts which we call  the unlocked regime (where the mean-field
solutions are correct) and the oscillating regime (where the
mean-field solutions fail). 
In the unlocked regime, there is one and only one time-independent 
stationary state and there is only one stationary value of the 
order parameter $c=<\cos x>$ which is always stable. In this regime, 
the reduced stationary probability density $P_{st}(x) \ne 0$ for any 
$x$. It means that with non-zero probability the phases of the oscillators 
can take any value of $x$ and  oscillators are not synchronized. 
In the oscillating regime, the only 
stationary state is 
temporarily oscillating state for which $\lim_{t \to \infty} P(x, t)$ 
is time-periodic and the order parameter $c = c(t)$ is time-periodic. 
From previously discussed  results it follows that this regime is bounded 
from the right, i.e. if $K > K_0$ then this regime disappears. 
The critical value $K_0$ can be determined by Eq. (\ref{c4}) 
from the condition that it possesses the double root $c_1 = c_2$.      
 The oscillating
regime is presented in Fig. \ref{fig6}, where we show evolution of two
distributions, the full density ${\hat P} (x, t)$ normalized on the
interval $\left(-\infty, \infty\right)$ and the reduced density $P(x,
t)$ normalized on one period.  In the latter case, the density
oscillates between the distribution of one maximum around $\pi$ (it
corresponds to the maximum of the local potential $-\cos x)$ and the
distribution of two maxima around $0$ and $2\pi$ (it corresponds to
the minima of the local potential).

 In turn, the non-diffusive region is
divided into two other parts which we call  the locked and hysteretic 
regimes. In the locked regime, only one steady-state solution exists. 
 In this regime, there are intervals of $x$ for which the reduced stationary 
probability density  $P_{st}(x)= 0$ and the phases of the oscillators 
 are locked in these intervals. It is an effect of interaction  and 
corresponds to the synchronization of oscillators (let us remember that 
in the case on non-interacting oscillators, the diffusive regime is realized 
in which the  phases can take arbitrary values). The synchronization 
is stronger if the support of  $P_{st}(x)$ is smaller. 
In this regime, there is only one mean-field value of $c = < \cos x>$ 
which is always stable. 
The so-called MF hysteretic regime is defined in the following way: 
There are three mean-field stationary values 
$c_1 > c_2 > c_3 > 0$ of the order parameter. The solutions $c_1$ and $c_3$ 
 are stable while $c_2$ is unstable.  
However, in this regime only one mean-field solution $c_1$ is realized,
which lies on the upper branch of the mean-field hysteresis,  
cf. the case $\tau =2$ for $K>4$ in Fig. \ref{fig4}.  
There is also  a regime of bistability. As in the previous case,
there are three mean-field stationary values 
$c_1 > c_2 > c_3 > 0$ of the order parameter. But now, two  stable solutions 
$c_1$ and $c_3$ can be realized what is demonstrated in Fig. \ref{fig7}. 
The upper state $c_1 > c_3$ corresponds to the locked regime, 
$a < 1+Kc_1(\tau)$  and 
the lower state $c_3$ corresponds to the unlocked regime, $a > 1+Kc_3(\tau)$, 
cf. Fig. \ref{fig4}, the case $K=2.94$ and $\tau =0.5$.     
It is also
instructive to see how the probability distributions $P(x, t)$ or
${\hat P}(x, t)$ evolve in time approaching the long time limit.  In
Fig. \ref{fig7}, the evolution of the density $P(x, t)$ is shown for
the values of parameters chosen from the bistability regime of the phase
diagram, i.e., when two stable stationary solutions exist. One can
observe that in dependence of the microscopic initial conditions the
system evolve either to the diffusive stationary state or to the
non-diffusive locked stationary state.  In two cases, the macroscopic
state, i.e., the initial probability density of oscillators is the
same uniform distribution. The microscopic state, i.e., initial
positions of all ``particles'' and realizations of noises are
different, it determines evolution of $P(x, t)$.
 For illustrating animations of the
time evolution we refer to our webpage \cite{web_mk}.

\begin{figure}[htbp]
  \begin{center}
    \epsfig{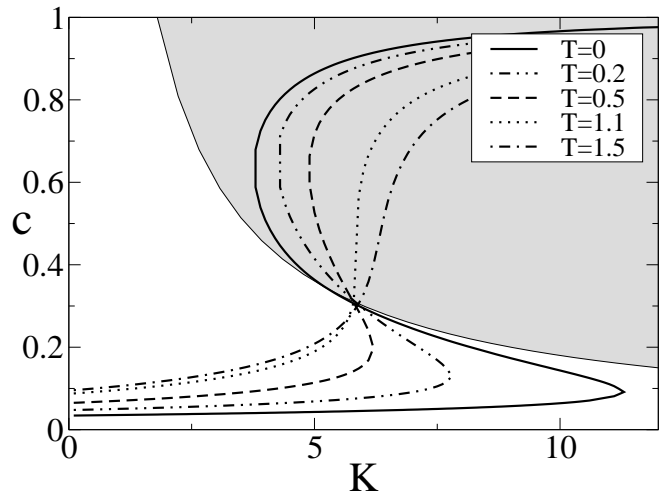}
  \end{center}              
  \caption{
    The order parameter $c$ versus coupling strength $K$ for selected
    values of temperature $T$. The increase of $T$ decreases the
    region of hysteresis.  Remaining parameters are $\tau=2.0$ and
    $a=2.8$.}
\label{fig8}
\end{figure}

The influence of temperature is depicted in Fig. \ref{fig8} (only
the mean-field case is shown). On the basis of these results, one may
conclude that the increase of thermal fluctuations acts like the
decrease of correlation time $\tau$ of nonequilibrium fluctuations.
The hysteretic region in $K$ is reduced as temperature grows. In
particular in Fig.  \ref{fig8} we see that for $T=1.5$ the mean-field
problem has got only a single solution.

\section{Summary}
\label{sec:summary} 
In this paper we have investigated the equilibrium and nonequilibrium
system of coupled phase oscillators. In fact, it can be any abstract
model of interacting particles in spatially periodic structure with a
periodic global interaction (e.g. interacting Brownian motors 
\cite{rei2,wio}).  The
equilibrium system defined by eq.  (\ref{pha1}) is a special case of
models considered in the literature.  Nevertheless, to the best of our
knowledge, the state equation (\ref{c}) has not been presented. We pay
attention to the subtle stability problem which sometimes is treated
superficially \cite{wio}.  Properties of the nonequilibrium system
(\ref{pha2}) are naturally much more interesting. The phase diagram
consists of five parts and  cannot be fully obtained from the
mean-field approach.  The non-mean-field regime is the oscillating
regime, which has been detected by use of the Monte-Carlo simulations
and by analyzing fluctuations of the order parameter $c=\langle \cos
x\rangle$.  The next interesting finding is that although the
non-interacting system is in the diffusive regime, the interaction can
move the system to the non-diffusive regime and then ``particles'' are
confined in valleys of the potential (of course it is exact when
temperature $T=0$).  It means that effectively, for the one-particle
dynamics, the barrier height $2\left( 1+Kc\right)$ of the local
potential is magnified and nonequilibrium fluctuations of amplitude
$a$ are not able to induce transitions over barrier.
 
All the results so far refer to the simple reflection-symmetric local
potential $-\cos x$.  If we add the higher order harmonics, e.g.
$\cos 2x$, the potential is still symmetric.  However, behavior of the
system can then be radically different because the second order
parameter $s=\langle \sin x\rangle \ne 0$.  New phenomena such as the
symmetry breaking, phase transitions and noise-induced transport can
occur in the system.  The paper on this subject will be published
elsewhere.

\section*{Acknowledgment}
\label{sec:ack} 
The work supported by Komitet Bada\'n Naukowych through the Grant No.
2 P03B 160 17 and The Foundation for Polish Science.

\bibliographystyle{unsrt}

\end{document}